\documentclass[journal=jacsat,manuscript=article]{achemso}

\usepackage{chemformula} 
\usepackage[T1]{fontenc} 
\usepackage{amsmath}
\newcommand{\vect}[1]{\mathbf{#1}}

\author{Derek M. Kita}
\affiliation{Department of Materials Science \& Engineering,\\ Massachusetts Institute of Technology, Cambridge, MA 02139}
\altaffiliation{Materials Research Laboratory,\\ Massachusetts Institute of Technology, Cambridge, MA 02139}
\alsoaffiliation{These authors contributed equally to this work.}
\email{dkita@mit.edu}
\author{J\'{e}r\^{o}me Michon}
\affiliation{Department of Materials Science \& Engineering,\\ Massachusetts Institute of Technology, Cambridge, MA 02139}
\altaffiliation{Materials Research Laboratory,\\ Massachusetts Institute of Technology, Cambridge, MA 02139}
\alsoaffiliation{These authors contributed equally to this work.}
\email{jmichon@mit.edu}
\author{Juejun Hu}
\affiliation{Department of Materials Science \& Engineering,\\ Massachusetts Institute of Technology, Cambridge, MA 02139}
\altaffiliation{Materials Research Laboratory,\\ Massachusetts Institute of Technology, Cambridge, MA 02139}

\title{A packaged, fiber-coupled waveguide-enhanced Raman spectroscopic sensor}

\begin{document}







\begin{abstract}
Waveguide-enhanced Raman spectroscopy (WERS) is a promising technique for sensitive and selective detection of chemicals in a compact chip-scale platform. Coupling light on and off the sensor chip with fibers however presents challenges because of the fluorescence and Raman background generated by the pump light in the fibers; as a result all WERS demonstrations to date have used free-space coupling via lenses. We report a packaged, fiber-bonded WERS chip that filters the background on-chip through collection of the backscattered Raman light. The packaged sensor is integrated in a ruggedized flow cell for reliable measurement over arbitrary time periods. We also derive the figures of merit for WERS sensing with the backscattered Raman signal and compare waveguide geometries with respect to their filtering performance and signal to noise ratio.
\end{abstract}

\section{Introduction}
Integrated waveguide Raman sensors have attracted significant interest notably for their compact size and large interaction volume that scales with the waveguide length. High-index contrast dielectric waveguides enable efficient excitation via evanescent modal fields. Raman scattered light is coupled back into these waveguides and sent to a spectrometer for chemical analysis. Prior work has demonstrated the validity and promises of this sensing technique using various waveguide material platforms including silicon nitride (Si$_3$N$_4$), alumina (Al$_2$O$_3$), Ta$_2$O$_5$, and TiO$_2$~\cite{Dhakal2014, Dhakal2016a, Evans2016, Holmstrom2016, Raza2018, Tyndall2018a, Zhao2018a, Coucheron2019, Raza2019}. Unfortunately, not all WERS system components can be easily integrated on a single chip, since WERS requires high-power monochromatic light sources at visible or near-infrared wavelengths, high-extinction ratio filters, and sensitive detectors. As a result, waveguide-enhanced Raman sensors are currently only functional when used in conjunction with off-chip components. Demonstrations of WERS to date have all utilized free-space, high-numerical-aperture (high-NA) objectives to couple light on and off chip. While this technique provides good coupling efficiencies, the need for expensive alignment stages and vibration sensitivity of these setups prevents them from being practical for field testing. Coupling light to Raman chips with optical fibers is the straightforward solution to this problem, yet the Raman pump signal generates unwanted fluorescence and Raman background in the input and output fibers~\cite{Tyndall2018}.  

In this work, we describe a scheme for collecting backscattered light using a simple on-chip beamsplitter and two long spiral waveguide sections. This method has two advantages over prior demonstrations that involve sensing of forward-scattered Raman light (despite imposing additional 3\,dB loss from the beamsplitter): (1) the pump and fiber Raman/fluorescence background predominantly propagate in the forward direction whereas half of the Raman signal coupled into the waveguide is backward-propagating (with the same intensity as the forward-propagating signal), giving rise to improved signal-to-noise and signal-to-background ratios (SNR and SBR); and (2) there is no waveguide length limit proportional to 1/$\alpha_s$ ($\alpha_s$ being the scattering losses) -- rather the waveguide sensing region can be made as long as possible to maximize the signal without loss penalty.

In Section~\ref{sec:experiment}, we first present a silicon nitride photonic integrated circuit (PIC) with exposed sensing windows that operates based on this technique. Fibers are bonded to both the input and output facets of the chip and a custom, ruggedized enclosure is machined to allow for in-line measurements of liquid or gaseous chemicals, as shown in Section~\ref{sec:measurement}. We then analyze the efficiency of this technique for filtering out pump signal and fiber background, and examine the effects of this collection scheme on SNR, as described in Section~\ref{sec:background}. Finally, in Section~\ref{sec:future}, we describe several techniques for improving the performance of future fiber-coupled waveguide Raman sensors.

\section{Chip design and fabrication}\label{sec:experiment}

The silicon nitride photonic sensor was made using a custom fabrication process at the MIT Microsystems Technology Laboratory. The final layer stack of the device consists of a 3\,$\mu$m bottom thermal oxide, a 200\,nm thick LPCVD silicon nitride waveguide layer, and a 2\,$\mu$m tetraethyl orthosilicate (TEOS) SiO$_2$ top cladding. The ridge waveguide is partially etched such that $\sim$35\,nm of silicon nitride remains.  This thin layer acts as a wet etch stop when opening sensing windows with a buffered oxide etch. A deep oxide etch step (5\,$\mu$m depth) defines facets for edge coupling a flat cleaved single-mode fiber (Nufern 780-HP) to the waveguides, and a deep reactive ion etch step (150\,$\mu$m depth) removes enough substrate so the fiber can access the facet.

\begin{figure}[bp]
\centering
\includegraphics[width=\linewidth, keepaspectratio]{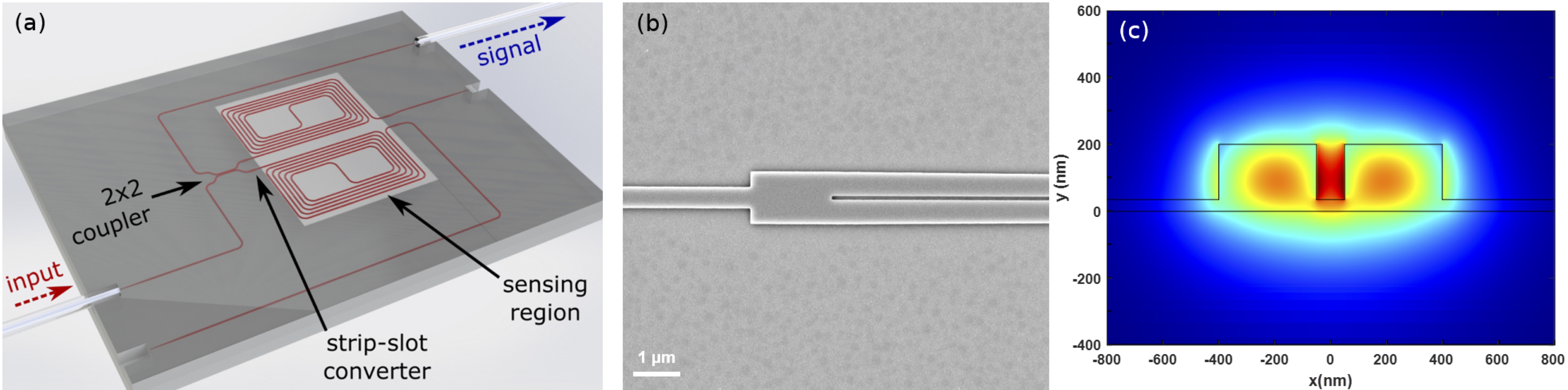}
\caption{(a) Rendering of the photonic chip design fabricated in this work. A $2\times2$ directional coupler splits light into two waveguide channels, a MMI converter transforms the strip to a slot mode, and then in an exposed sensing region Raman scattered light is coupled into the backward-propagating waveguide mode. Components are not to scale. (b) SEM image of the MMI strip-to-slot converter. (c) Slot waveguide cross-section overlaid with the simulated modal intensity for the fundamental TE mode.  }
\label{fig:layout}
\end{figure}

On the chip (see Fig.~\ref{fig:layout}(a)), 100\,$\mu$m long inverse spot-size converters with 75\,nm wide tips expand the waveguide mode to allow for more efficient fiber coupling (with a theoretical $-3.0$\,dB/facet insertion loss, and a measured loss of $-7.2$\,dB/facet). The single-mode waveguides are 500\,nm wide, and a $2\times2$ directional coupler (20\,$\mu$m long, 0.4\,$\mu$m gap) is used to split light into two arms. A multi-mode interferometer (MMI) with 2.7\,$\mu$m $\times$ 1.5\,$\mu$m dimensions (Fig.~\ref{fig:layout}(b)) is used to convert the TE polarized strip mode to a TE polarized slot mode \cite{Deng2016a}. The slot waveguide consists of two 350\,nm rails and a 100\,nm slot gap (Fig.~\ref{fig:layout}(c)) with a total waveguide width of 800 nm. These waveguides then enter the spiral waveguide sensing region which is 8\,cm long and consists of 100\,$\mu$m radius bends. Backward-propagating Raman generated light travels back to the $2\times2$ coupler where approximately half of the light is carried to an output port at the opposite edge of the chip as the input.

To bond optical fibers to the edge of the chip, we begin by mounting each fiber into a glass V-groove block (OZ Optics) such that the fiber tip extends $\sim$0.5\,mm beyond the glass block. We then bond the fiber and V-groove block with UV curable epoxy. Next, the fibers and the chip are each mounted in custom 3D-printed arms that attach to 5-axis positioning stages for both input and output alignment. Our process for applying epoxy to each fiber consists of first touching the tip of each fiber to a small droplet of epoxy under a microscope and then gently retracting the fiber so that only a very small amount of transparent UV-curable epoxy wets the flat fiber tip area. This process ensures that a minimal volume of epoxy is used during fiber-to-chip bonding, since volume changes during curing (resulting in sub-micron level displacements) can have dramatic, detrimental effects on the coupling efficiency. Once the fibers on both ends are aligned and the power transmitted through the waveguides is optimized, the epoxy is cured. Next, larger volumes of epoxy are applied between the glass block and the silicon nitride photonic chip for mechanical stability.

\begin{figure}[h]
\centering
\includegraphics[width=.7\linewidth, keepaspectratio]{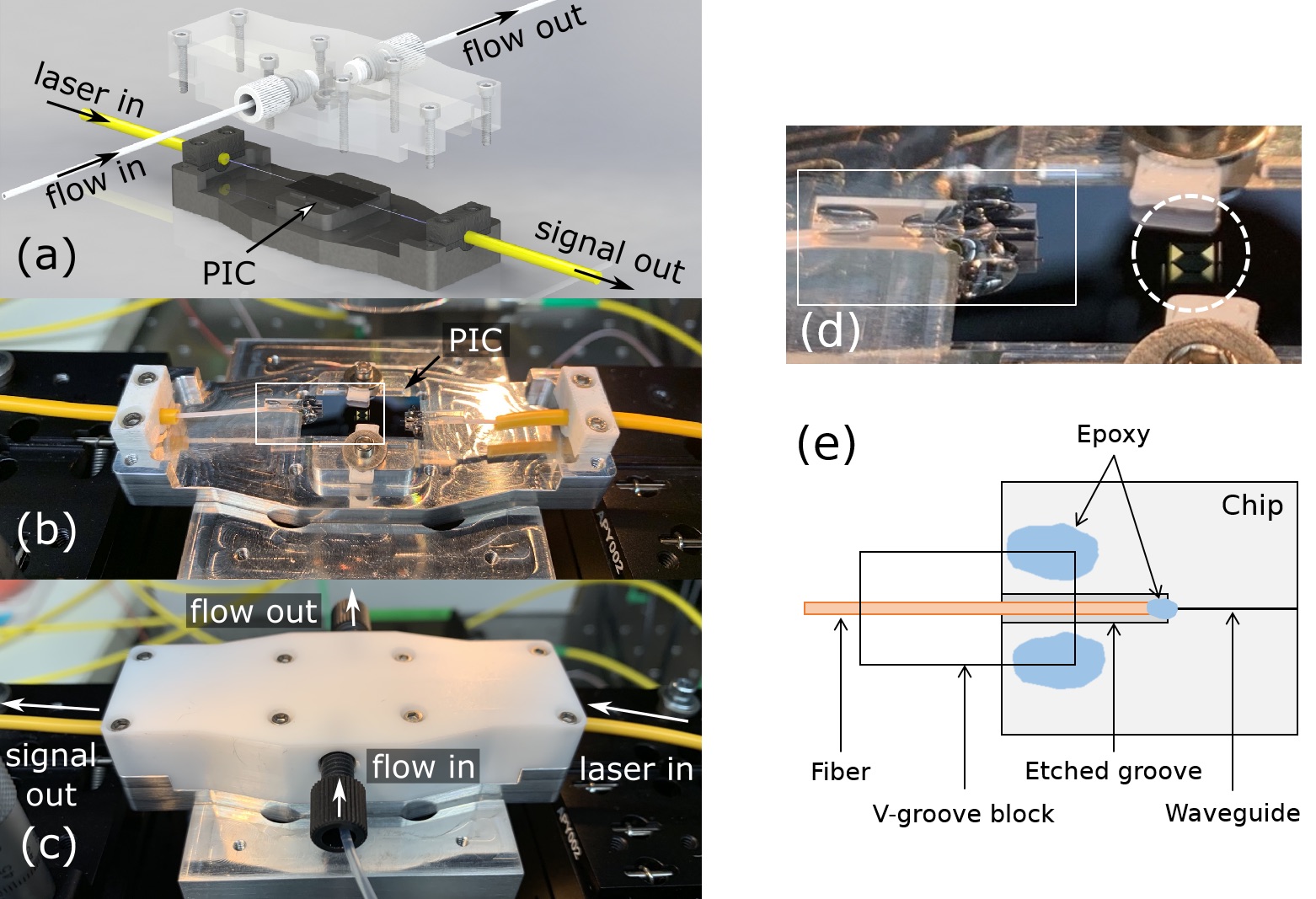}
\caption{(a) Rendering of the chip enclosure with optical fibers affixed to the chip and a flow cell that lowers onto the surface of the chip. The top case and flow channels are machined from PTFE, a rubber Kalrez$^\text{\textregistered}$ O-ring separates the chip from the top case, and the bottom fixture is machined from aluminum. (b) Photograph of the fiber-coupled, packaged Raman sensing chip with fibers glued to the edge of the chip after being mounted in separate glass V-groove chips (for mounting purposes). (c) The packaged chip with the top PTFE cell secured above the chip. (d) Zoomed-in top view of the area boxed in white on (b). The position of the O-ring is shown by the dashed white circle (the side clamps are removed prior to O-ring positioning). (e) Top-view schematics of the fiber-bonding region (boxed in white on (d)). Components are not to scale. The glass V-groove block is bonded to the chip with epoxy at the tip of the fiber in the etched groove, and with larger volumes of cured epoxy on the sides of the block for improved robustness.}
\label{fig:cad_and_image}
\end{figure}

In order to use the chip for sensing a variety of compounds, including organic solvents like isopropanol or acetone that would readily dissolve the epoxy at the edge of chips, we designed a ruggedized flow cell as shown in Fig.~\ref{fig:cad_and_image}. The cell consists of a machined aluminum bottom component that the photonic sensor chip sits on top of, a polytetrafluoroethylene (PTFE) top component with drilled holes that form the flow cell, and a rubber Kalrez$^\text{\textregistered}$  O-ring that separates the chip from the top PTFE component. The top and bottom pieces screw tightly together to put pressure on the O-ring and seal off the sensing region. The PTFE top component and rubber O-ring are both chemically resistant to a wide range of organic solvents and acids and can withstand high temperatures, enabling sensing of liquids in harsh environments (e.g. in flow-chemistry reactions). Using this ruggedized flow cell with our fiber-packaged waveguide-enhanced Raman sensors, we are able to measure a variety of chemical species with arbitrary integration times and without the need for any optical alignment.

\section{Measurements}\label{sec:measurement}
The optical measurement setup consists of a wavelength-stabilized single mode fiber coupled laser operating at $\lambda=808$\,nm (QPhotonics). A fiber-integrated narrow bandpass filter centered at $\sim$810\,nm is used to remove amplified spontaneous emission from the signal, and polarization control paddles before the chip are used to couple light into either the TE or TM waveguide mode. After the chip, light travels to a fiber bench with a notch filter to prevent any remaining pump light at 808\,nm from continuing to the spectrometer. The Raman signal is sent via fiber to a compact commercial spectrometer with a cooled CCD (AvaSpec-HERO, Avantes). The total fiber length in the setup is around 2\,m. Although most of the pump light is removed on-chip by the forward-propagating waveguide mode, finite reflections on the chip necessitate the additional notch filter. The optical components that contribute the most reflection are the strip-to-slot waveguide mode converter, the transition from the waveguide being top clad with oxide to top clad with analyte, and the backscattered light from waveguide roughness in the sensing region.

\begin{figure}[h]
\centering
\includegraphics[width=\textwidth, keepaspectratio]{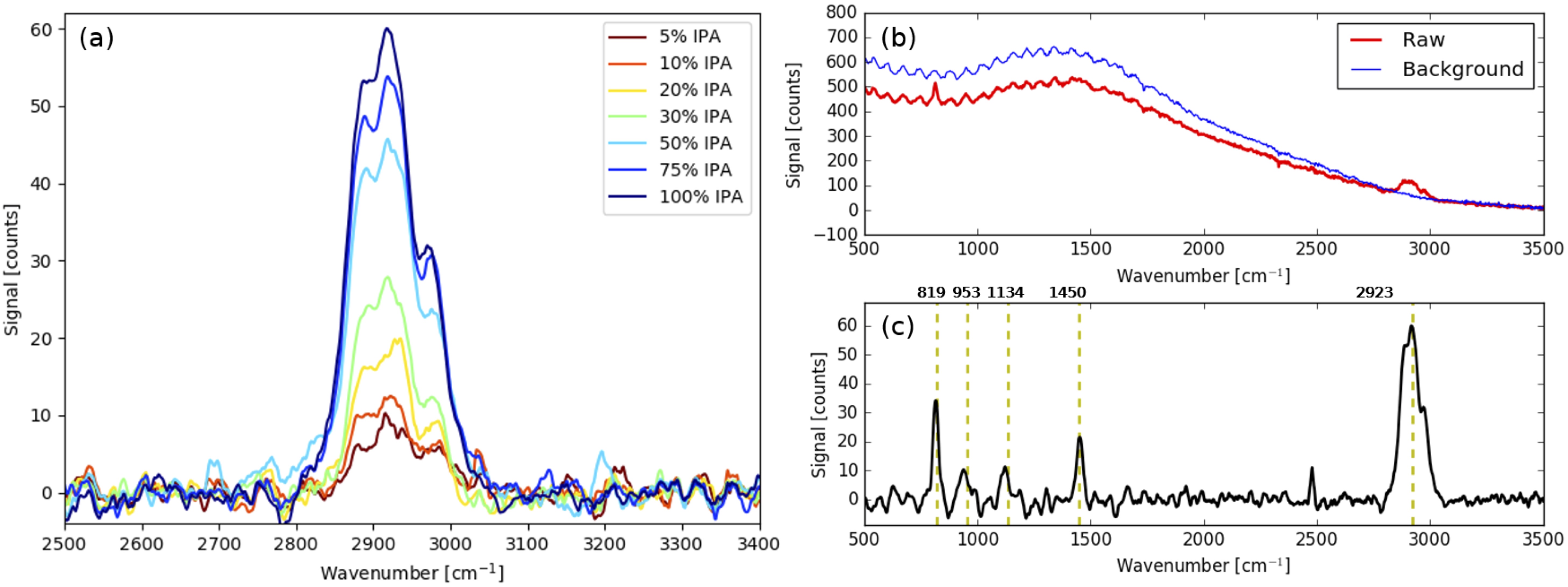}
\caption{(a) Measured Raman spectra of mixtures of isopropyl alcohol (IPA) and water at different weight fractions, after background subtraction and baseline correction. Each spectra is the result of $15\times$ 60-second integrations. The standard deviation on each wavelength measurement is approximately 2.9 counts. (b) Raw and background spectra of the 100\% IPA measurement. The low-frequency drift between the two spectra is eliminated with a subsequent baseline correction. (c) Full spectrum of the 100\% IPA measurement after background and baseline correction, with the characteristic peaks of IPA.}
\label{fig:spectra}
\end{figure}

Using a packaged chip with 800\,nm wide slot waveguides with a 100\,nm gap, Raman spectra were obtained for varying concentations of isopropyl alcohol (IPA) in water, as shown in Fig.~\ref{fig:spectra}. A separate measurement with only deionized water above the chip was used to subtract any background fluorescence and wavelength dependence of light coming off the chip since reflection at several on-chip components generated weak Fabry-P\'{e}rot fringes (Fig.~\ref{fig:spectra}(b)). Due to the wavelength dependence of the directional coupler, we also saw greater signal collection for Raman light at the peaks around 2923\,cm$^{-1}$ than at peaks between 819-1450\,cm$^{-1}$. For this device, the SBR (ratio of Raman signal from IPA above the chip to background signal measured at the bottom of the peak, without any baseline subtraction) was $58:62$ (counts) for the 2923\,cm$^{-1}$ peak with 100\% IPA. In order to identify the source of the background signal and to extract the additional pump light rejection when collecting backscattered Raman signal, we measured the forward-scattered Raman signal from a separate 5\,mm-long straight slot waveguide. In this configuration, the strong background fluorescence and Raman generated in the fiber (with an amplitude of $898$ counts) prevented direct identification of the IPA peaks. Only when subtracting a baseline measurement performed with no analyte could the Raman peaks at 2923\,cm$^{-1}$ be identified, with an amplitude of $143$ counts. As we will discuss in more details in Section~\ref{sec:background}, these SBR values indicate that the background signal originates mostly from the fiber, and allow us to extract an effective on-chip pump rejection ratio of $9.6$\,dB as a result of collecting the backreflected light. This agrees well with our calculations from Section~\ref{sec:background} as it is slightly lower than the theoretical maximum of $\sim$12\,dB for a 100\,nm-gap TE slot waveguide (see Fig.~\ref{fig:performance}(d)).

\section{Analysis of backscattered Raman signal}\label{sec:background}

The useful signal in the backscattered sensing scheme is the power of backward-propagating Raman light collected at the start of the sensing region ($x=0$). 
Setting the powers to zero as $x \rightarrow \infty$, the ratio of Raman power to input power is given by \cite{Raza2019}:
\begin{equation}
P_\text{Raman} = \frac{\beta_\text{Raman}}{2\alpha_s} P_f(0)\,,
\label{eq:P_Raman}
\end{equation}
where $\beta_\text{Raman} = \rho_\text{analyte} \sigma_\text{analyte} \cdot \lambda^2 n_g^2\iint_\text{clad}|\vect{E}|^4/\left(\iint_\text{cs} \epsilon|\vect{E}|^2\right)^2$ is the generation rate of the Raman signal by the the analyte, and $\alpha_s$ is the scattering loss per unit length. The collected Raman signal is similar to that derived for the forward-scattered light collection scheme, for which $P_\text{Raman} = (\beta_\text{Raman}/\alpha_s) e^{-1}P_f(0)$ \cite{Kita2018a}. 

An advantage of collecting backscattered Raman light is that the fiber fluorescence generated by the input fiber and the pump light (that would generate fiber fluorescence in the output fiber) should couple only weakly to the backward-propagating mode. The corresponding rejection ratio is thus determined by the amount of reflected (backward-propagating) pump light normalized by the input (forward-propagating) pump light. Assuming very low-loss components with adiabatic mode transformations, the dominant source of back reflection is scattering from sidewall roughness $\alpha_r$. This, along with the scattering loss (power scattered only into the far field) $\alpha_l$, determines the distance that light propagates into the sensing region. The rate equations determining the pump power in the forward-propagating mode $P_f$ and the pump power in the backward-propagating mode $P_b$ are given by:
\begin{align}
\frac{dP_f(x)}{dx} &= -\alpha_l P_f(x) - \alpha_r P_f(x) + \alpha_r P_b(x) \,,\\
\frac{dP_b(x)}{dx} &= \alpha_l P_b(x) + \alpha_r P_b(x) - \alpha_r P_f(x) \,.
\end{align}
Solving these coupled equations with the boundary conditions $\lim_{x \rightarrow \infty} P_f(x)=0$ and $\lim_{x \rightarrow \infty} P_b(x)=0$ yields a relationship between the input power $P_f(0)$ and the output reflected power $P_b(0)$:
\begin{equation}
P_b(0) = \Big[1 + f - \sqrt{2f + f^2}\Big] P_f(0) \equiv \eta P_f(0) \,,
\label{eq:P_b}
\end{equation}
where $f = \alpha_l/\alpha_r$ is the ratio of scattered power loss to reflected power, and $\eta$ quantifies the rejection ratio.

We use these expressions to calculate the SNR and SBR for the backscattered configuration. Only taking into account parameters affected by the waveguide geometry and component design, we find in Appendix A that:
\begin{align}
& \text{SNR} \sim \sqrt{\gamma_\text{dc}} \dfrac{\beta_\text{analyte}/2\alpha_s}{\sqrt{\beta_\text{analyte}/2\alpha_s + \beta_\text{core}/2\alpha_s + \eta\beta_\text{fibers} L_\text{fibers}}} \,, \\
& \text{SBR} \sim \dfrac{\beta_\text{analyte}/2\alpha_s}{\beta_\text{core}2\alpha_s + \eta\beta_\text{fibers} L_\text{fibers}} \,,
\end{align}
where $L_\text{fibers}$ is the combined length of the input and output fibers, and $\beta_\text{core}$ (resp. $\beta_\text{fibers}$) is the generation rate of all processes happening in the waveguide core (resp. in the fibers) that contribute to the background, such as Raman scattering and fluorescence.

Comparing these expressions to the experimental results from Section~\ref{sec:measurement} enable us to identify the relative contributions of the analyte, waveguide core, and fibers, to the total power. First, we calculate the number of counts due to the waveguide core, which scales with the signal from the analyte via a factor $\beta_\text{core}/\beta_\text{analyte}$. In this we assume that the sensing region of the waveguide is much longer than the oxide-clad, non-sensing region where the core alone generates a signal. Using values for $\rho\sigma$ from Refs.~\cite{Dhakal2017,Colles1972}, the $1/\lambda_\text{pump}^4$ dependence of the cross-section, and mode profile simulations to calculate the field integrals, we find that $\beta_\text{pure IPA} = 2.16 \times 10^{-9}\,\text{cm}^{-1}$ and $\beta_\text{core} = 0.54 \times 10^{-9}\,\text{cm}^{-1}$ for a Stokes shift of 2923\,cm$^{-1}$ with $\lambda_\text{pump} = 808$\,nm. The IPA signal is thus $\sim4\times$ higher than the signal from the waveguide core. For our sample in the forward-collection configuration, this amounts to 36 background counts due to the core, out of 898 total background counts. For the backward-collection sample, the core contributes 15 counts out of 62 total background counts. In both cases, we conclude that the background mostly originates from the fibers. While the background still is not dominated by the signal from the waveguide core, which is the ultimate limit for the background signal \cite{LeThomas2018}, the fiber contribution to the background is nonetheless greatly reduced by the use of the backward-collection scheme. After accounting for the loss due to the directional coupler, the difference in fibers background counts between the two configurations indicates a 9.6\,dB rejection ratio from our measurement. 

\begin{figure}[h!]
\centering
\includegraphics[width=0.8\linewidth, keepaspectratio]{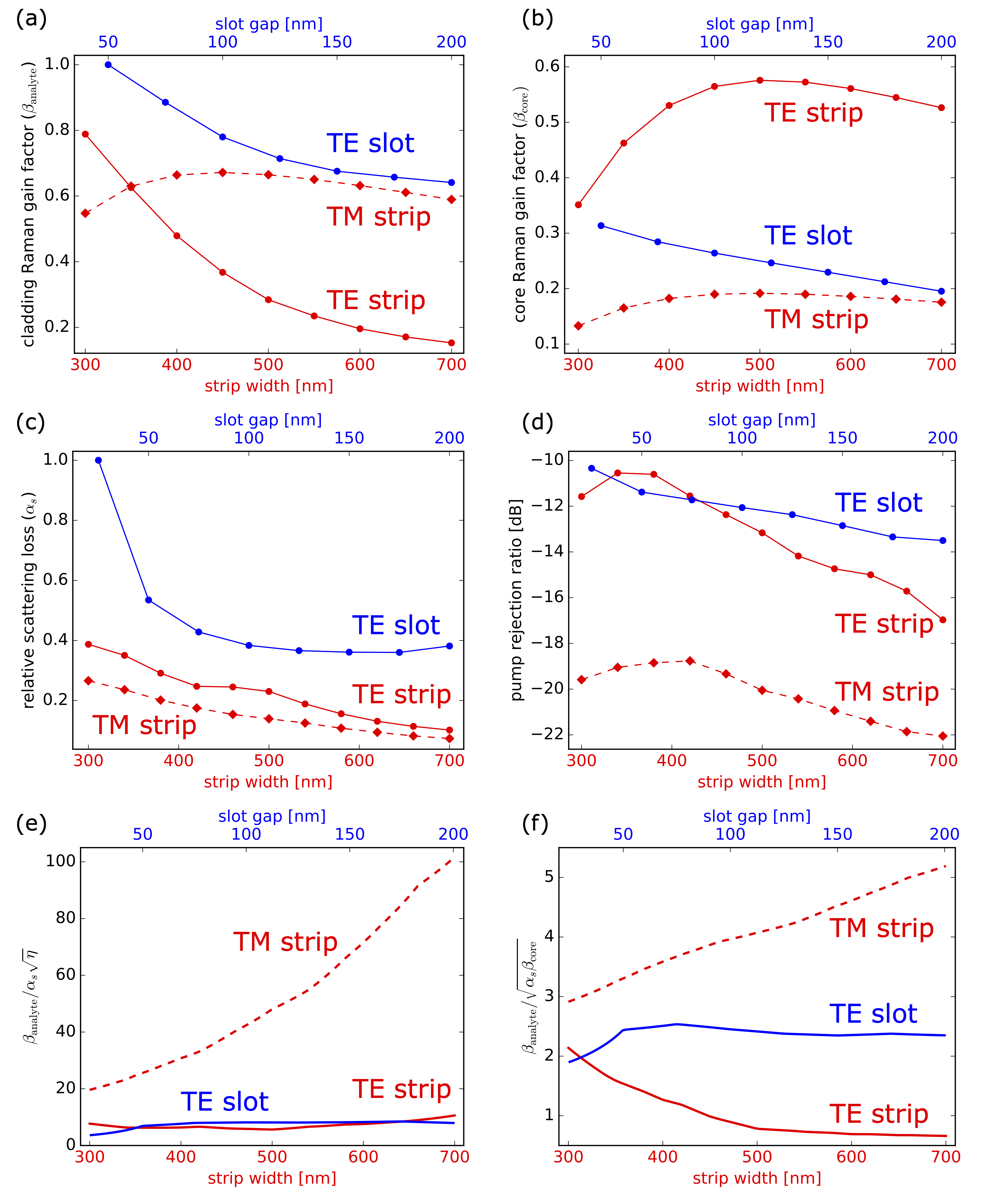}
\caption{Numerically computed values of (a) the relative cladding Raman gain coefficients (normalized by the 50\,nm TE slot value) computed via field integrals over the top clad (analyte sensing) region, assuming water as the solvent; (b) the relative core Raman gain coefficients computed via field integrals over the waveguide core region; (c) the scattering loss $\alpha_s$ computed via the volume-current method; (d) the pump rejection ratio $10 \log_{10}(\eta)$, where $\eta$ is given by Eq.~(\ref{eq:P_b}); (e) the fiber-limited backscattered Raman sensing FOM $\beta_\text{analyte}/\alpha_s\sqrt{\eta}$ with data interpolated from (a), (c), and (d); and (f) the core-limited backscattered Raman sensing FOM $\beta_\text{analyte}/\sqrt{\alpha_s\beta_\text{core}}$ with data interpolated from (a), (b), and (c).  Slot waveguides are all 800\,nm wide.}
\label{fig:performance}
\end{figure}

Being in the fiber-limited background regime also allows us to simplify the expressions of the SNR and SBR. In particular, the figure of merit (FOM) to compare different waveguide geometries for high-sensitivity sensing in the backscattered scheme is given by the SNR in the low-concentration limit, which depends on the waveguide geometry as:
\begin{equation}
\text{SNR} \sim \dfrac{\beta_\text{analyte}}{\alpha_s\sqrt{\eta}} \,.
\end{equation}
We can also quantify the performance improvement of the backward-collection configuration over the forward-collection one (for a fiber-limited background):
\begin{equation}
\dfrac{\text{SNR}_\text{b}}{\text{SNR}_\text{f}} = \sqrt{\dfrac{\gamma_\text{dc} e}{4 \eta}} \,,
\end{equation}
where the subscript ``$b$'' (resp. ``$f$'') denotes the backward-collection (resp. forward-) configuration. Assuming a perfect beamsplitter such that $\gamma_\text{dc} = -3$\,dB, the backscattered scheme yields an improved SNR as long as $\eta \leq -4.6$\,dB, which is easily realized with common waveguide geometries (see Fig.~\ref{fig:performance}(d)). Finally, for very high on-chip rejection, the background signal due to the fibers becomes negligible and the waveguide core becomes the major source of background signal. The SNR then scales as:
\begin{equation}
\text{SNR} \sim \dfrac{\beta_\text{analyte}}{\sqrt{\alpha_s\beta_\text{core}}} \,.
\end{equation}
As expected, a similar scaling appears for lens-coupled WERS \cite{Dhakal2017} -- with the additional $1/\sqrt{\alpha_s}$ factor here accounting for the sensing waveguide length. The only difference between the backscattered fiber-coupled and lens-coupled SNRs then resides in the edge-coupling efficiency and the coupler loss. 

We then proceed to compare different waveguide geometries for backscattered WERS. Since $\eta$ depends only on the ratio $f$, and $\alpha_r$ and $\alpha_l$ are both proportional to the net waveguide scattering loss $\alpha_s$, the rejection ratio is a function of the local density of states at the scattering center~\cite{Oskooi2013}. Assuming that waveguide sidewall roughness is the dominant scattering mechanism and the roughness correlation length is significantly larger than the RMS roughness amplitude, we use the volume-current method \cite{Kita2018a} to numerically compute $\alpha_l$, $\alpha_r$, and $f$ for a variety of TE and TM strip waveguide modes and 800\,nm wide slot waveguides with varying slot gaps at $\lambda=808$\,nm. The results of this analysis are shown in Fig.~\ref{fig:performance}, and suggest that TM strip waveguides offer the best tradeoff between Raman gain, scattering loss, and pump rejection (Fig.~\ref{fig:performance}(e)) or waveguide core background (Fig.~\ref{fig:performance}(f)). Despite enhanced field confinement in the top cladding for narrow gap slot waveguides, the field enhancement in the relatively low index contrast structure ($n_\text{clad}=1.329$, $n_{\text{Si}_3\text{N}_4}=2.02$) is not enough to overcome the higher losses as a result of the additional sidewall interface. In addition, TM strip waveguides were observed to have nearly $10\times$ better pump rejection ratios than TE structures because scattered power is more likely to couple back into the fundamental TE mode than the TM mode.

\section{Conclusion}\label{sec:future}
Using a fiber-coupled silicon nitride WERS sensor designed to collect only the backreflected Raman light, we demonstrate a functional, robust sensing device that circumvents the issue of fiber fluorescence/Raman background. The machined enclosure with integrated flow cell enabled measurements of IPA in water down to single-digit weight percentages. A key advantage of this technique is that it allows for up to $\sim20$\,dB of fiber background suppression and there is no fabrication-dependent optimal waveguide length for the spiral waveguide sensor. Our analysis of the SNR confirms the benefits of this collection scheme over the forward-collection configuration, and yields the FOM to consider when choosing a waveguide geometry. Furthermore, our volume-current method calculations suggests that narrow gap structures (like slot waveguides) may not yield better waveguide sensors because of their higher scattering losses, while TM strip waveguides see their lower gain coefficient compensated by low scattering losses and high rejection ratio.

A number of improvements to future device designs can significantly improve the signal-to-noise ratio for high-performance Raman sensing applications. First, using TM strip waveguides instead of TE slot waveguides would both increase the output Raman signal and more efficiently filter out the pump source, overall increasing the SNR. Second, improving the coupling efficiency to the waveguides with better inverse taper couplers and better packaging techniques would also yield an improved SNR. Third, eliminating all components that contribute to back-reflection would reduce the transmitted pump light and fiber background. This can be accomplished by eliminating the MMI-based strip-to-slot waveguide converter. Also, reflections at the interface between the buried waveguide and the exposed sensing waveguide can be reduced by letting the waveguide enter the sensing region at a shallow angle (rather than at a normal angle). Finally, replacing the directional coupler with either a broadband adiabatic coupler or wavelength-selective filters \cite{Tyndall2018, Nie2019, Oser2019} would improve the flatness of the received Raman signal and could eliminate the 3\,dB insertion loss penalty.

\appendix

\section*{Appendix A: Signal-to-noise and signal-to-background ratios}\label{sec:app_a}

The figure of merit to compare different Raman configurations and WERS waveguide geometries is the SNR, defined as, for a detector photon count following a Poisson distribution:
\begin{equation}
\text{SNR} = \dfrac{C_\text{total}-C_\text{background}}{\sqrt{C_\text{total}}} = \sqrt{\theta t_\text{int}} \dfrac{P_\text{total}-P_\text{background}}{\sqrt{P_\text{total}}}
\end{equation}
where $C_\text{total}$ (resp. $P_\text{total}$) is the total number of counts (resp. total power) at a given peak in the Raman spectrum, $C_\text{background}$ (resp. $P_\text{background}$) is the number of counts (resp. power) of the background near that peak \cite{Dhakal2016a}, $\theta$ is the detector's responsivity, and $t_\text{int}$ is the integration time. We will omit the responsivity and integration time in the following. The SBR is given by:
\begin{equation}
\text{SBR} = \dfrac{C_\text{total}-C_\text{background}}{C_\text{background}} = \dfrac{P_\text{total}-P_\text{background}}{P_\text{background}}
\end{equation}

For fiber-coupled WERS, the total power includes contributions from the analyte, the waveguide core, and the fibers:
\begin{equation}
P_\text{total} = P_\text{Raman,analyte} + P_\text{core} + P_\text{fibers} = P_\text{Raman,analyte} + P_\text{background}
\end{equation}
where $P_{\text{Raman,analyte}}$ is the Raman power generated by the analyte as light travels along the waveguide, and $P_{\text{core}}$ (resp. $P_{\text{fibers}}$) is the power generated in the core of the waveguide (resp. in the input and output fibers) by any processes such as Raman and fluorescence. The peak height is given by $P_{\text{Raman,analyte}}$ and thus $P_{\text{background}} = P_{\text{core}} + P_{\text{fibers}}$. The SNR and SBR are thus:
\begin{align}
& \text{SNR} = \dfrac{P_\text{Raman,analyte}}{\sqrt{P_\text{Raman,analyte} + P_\text{core} + P_\text{fibers}}} \\
& \text{SBR} = \dfrac{P_\text{Raman,analyte}}{P_\text{core} + P_\text{fibers}}
\end{align}
\\

\subsection*{A.1 Forward-scattered configuration}

In the forward-scattered configuration, any light traveling through the setup undergoes losses due to fiber-to-chip coupling and propagation to the sensing region ($\gamma_\text{edge}$), propagation through the sensing region (of length $L$), and propagation from the sensing region and chip-to-fiber coupling ($\approx \gamma_\text{edge}$). The different powers are given by \cite{Kita2018a,Dhakal2014}:  
\begin{align}
P_\text{Raman,analyte} & = \gamma_\text{edge}^2 e^{-\alpha_s L} P_\text{pump} \times \beta_\text{analyte} L \\
P_\text{core} & = \gamma_\text{edge}^2 e^{-\alpha_s L} P_\text{pump} \times \beta_\text{core} L  \\
P_\text{fibers} & = \gamma_\text{edge}^2 e^{-\alpha_s L} P_\text{pump} \times \beta_\text{fibers} L_\text{fibers}
\end{align}
with $P_\text{pump}$ the pump power in the input fiber. In the optimal case, $L=1/\alpha_s$, with $\alpha_s$ the propagation losses, assumed to be the same at the pump and Stokes frequencies.

The expressions for the SNR and SBR in the forward-collection configuration are thus:
\begin{align}
& \text{SNR}_\text{f} = \sqrt{\gamma_\text{edge}^2 e^{-1} P_\text{pump}} \dfrac{\beta_\text{analyte}/\alpha_s}{\sqrt{\beta_\text{analyte}/\alpha_s + \beta_\text{core}/\alpha_s + \beta_\text{fibers} L_\text{fibers}}} \\
& \text{SBR}_\text{f} = \dfrac{\beta_\text{analyte}/\alpha_s}{\beta_\text{core}/\alpha_s + \beta_\text{fibers} L_\text{fibers}}
\label{eq:SBR_f}
\end{align}

\subsection*{A.2 Backscattered configuration}
In the backscattered configuration, the directional coupler incurs additional losses ($\gamma_\text{dc}$, with a minimum of 3\,dB). The expressions for the backward-propagating pump power and Raman power are given by Eq.~\ref{eq:P_Raman} and Eq.~\ref{eq:P_b}, yielding:
\begin{align}
P_\text{analyte} & = \gamma_\text{edge}^2 \gamma_\text{dc} P_\text{pump} \times  \dfrac{\beta_\text{analyte}}{2\alpha_s} \\
P_\text{core} & = \gamma_\text{edge}^2 \gamma_\text{dc} P_\text{pump} \times  \dfrac{\beta_\text{core}}{2\alpha_s}  \\
P_\text{fibers} & = \gamma_\text{edge}^2 \gamma_\text{dc} P_\text{pump} \times \eta \times \beta_\text{fibers} L_\text{fibers}
\end{align}

The expressions for the SNR and SBR in the backward-collection configuration are thus:
\begin{align}
& \text{SNR}_\text{b} = \sqrt{\gamma_\text{edge}^2 \gamma_\text{dc} P_\text{pump}} \dfrac{\beta_\text{analyte}/2\alpha_s}{\sqrt{\beta_\text{analyte}/2\alpha_s + \beta_\text{core}/2\alpha_s + \eta\beta_\text{fibers} L_\text{fibers}}} \\
& \text{SBR}_\text{b} = \dfrac{\beta_\text{analyte}/2\alpha_s}{\beta_\text{core}/2\alpha_s + \eta\beta_\text{fibers} L_\text{fibers}}
\label{eq:SBR_b}
\end{align}

\medskip

\subsection{Funding} 
Saks-Kavanaugh Foundation; MIT Deshpande Center for Technological Innovation.

\subsection{Acknowledgement}
The authors acknowledge fabrication support from the Microsystems Technology Laboratory at MIT.


\subsection{Disclosures} The authors declare no conflicts of interest.

\bibliography{packaged_raman_refs}

\end{document}